\documentclass[10pt,a4paper,reqno]{amsart}
\usepackage[utf8]{inputenc}
\usepackage{amsmath,hyperref}
\usepackage{amsfonts}
\usepackage{amssymb,amsthm,stmaryrd}
\usepackage{graphicx}
\usepackage{color}

\usepackage{tikz}

\newtheorem{remark}{Remark}
\usepackage[normalem]{ulem} 

\usepackage{geometry}
\usepackage{algorithm,algpseudocode}

\def\forward#1{\overrightarrow{ #1}}
\def\backward#1{\overleftarrow{ #1}}

\author{Bertrand Cloez,  Bénédicte Fontez, Eliel  Gonz\'{a}lez-Garc\'{i}a, Isabelle Sanchez}

\title[Kalman filter with impulse noised outliers ]{Kalman filter with impulse noised outliers : a robust sequential algorithm to filter data with a large number of outliers}


\begin{document}

\maketitle


\begin{abstract}
Impulse noised outliers are data points that differs significantly from other observations. They are generally removed from the data set through local regression or Kalman filter algorithm. However, these methods, or their generalizations, are not well suited when the number of outliers is of the same order as the number of low-noise data. In this article, we propose a new model for impulse noised outliers based on simple latent linear Gaussian processes as in the Kalman Filter. We present a fast forward-backward algorithm to filter and smooth sequential data and which also detect these outliers. We compare the robustness and efficiency of this algorithm with classical methods. Finally, we apply this method on a real data set from a Walk Over Weighing system admitting around $60\%$ of outliers. For this application, we further develop an (explicit) EM algorithm to calibrate some algorithm parameters.

\keywords

\tableofcontents
\end{abstract}

\section{Introduction}

Sequential data  with automatic records from sensors are the new era in most applied fields. The ubiquity of sensors leads to the emerging concept of novelty or anomaly detection \cite{PIMENTEL2014,Fisch2022} where outliers are considered as an interesting novelty or an expression of a relevant situation. A more common concern with automatic records is the need for pre-treatment to correct from flawed or erroneous measurements, for example when data were not recorded under normal conditions of use. In this case, noise are unwanted in the data and authors suggest to do a noise removal or a noise accommodation \cite{Chandola2009}. Both concepts can be addressed with a model describing the way data were generated, to provide predictive information and automatic detection of signal.

Assumption on the noise is required to build such a model.  For image or signal processing, a well known unwanted noise is the impulse noise\footnote{https://www.sciencedirect.com/topics/engineering/impulse-noise} (the salt and pepper in an image). This noise is modelized through fixed values or in a more general framework by some random values \cite{Maity2018}. 

Random-valued impulse noise can be defined in several ways. In this paper we use the following model:
\begin{equation}\label{eq:modelintro}
    Y = \left\{
    \begin{array}{ll}
       X + \varepsilon  &  \text{with probability } p\\
       O  & \text{with probability } 1-p
    \end{array}
    \right.
\end{equation}

where $Y$ is the observed value of signal of interest $X$. We assume that $\varepsilon$ has a Gaussian distribution and $O$ is the unwanted random outlier. Modelling the observation status $Z$ with a Bernoulli distribution, the model can be expressed as $Y=  Z (X+\varepsilon) + (1-Z) O$. We further assume that $X, Z, O$ and $\varepsilon$ are independent. This type of model will permit both to identify outliers and to reconstruct the longitudinal trajectory.  

Different algorithms allow an image reconstruction under impulsive noise and most of them \cite{Pardeshi2015} use spatial similarity (correlation) of adjacent pixels like the median filter. Here, we consider longitudinal data and assume similarly time-correlations through a Gaussian Markov model $(X_{k})_{k\geq 1}$. Consequently, the assumption of Equation~\eqref{eq:modelintro} is generalised with indices $k$. For every $k \geq 1$, variables $X_k, \varepsilon_k$ and $Z_k$ being non-observed, we are in a latent model. These assumptions lead to the problem of reconstructing the longitudinal path of a hidden Markov chain.

Recently, a popular approach to treat hidden Markov chain parameters is given by the \textit{sequential Monte-Carlo} (SMC) algorithm \cite{CMT07,DM13}. This algorithm is relatively efficient and permits to treat general non-linear and non-Gaussian models, for data sets including outliers, for the price of very high calculation time. Well adapted versions of SMC like \textit{Particle Swarm Optimization} (PSO) filter particles are a hot topic in the image recovery research \cite{Yao2011}. To avoid high computational complexity, a recent hybrid Kalman filter and PSO algorithm was proposed by \cite{Peng2020}.

Indeed, the Kalman filter is an efficient and widely used algorithm for estimating the state of a dynamical system, given noisy observed data \cite{G74,AM12,M82,K60,KB61, TTS07}. This is partly because it is the optimal estimator for linear Gaussian dynamical system, in the sense that it yields the smallest expected mean-squared error \cite{M76}. However, even for simple linear Gaussian dynamical system, this model breaks down in the presence of non Gaussian noise. Most authors use an additive noise with the use of Student (or PTV) distribution or a mix of Gaussian distributions to generate a heavy-tailed distribution to model the noise in addition to the Gaussian process as in \cite{SA71,MM77,MS89,SM94, WK96,W81, ANN11}. This adds robustness in the classical Kalman filter. Nevertheless, these approaches are substantially more difficult to implement and are computationally very demanding, especially for online updating or for coupling them with a likelihood optimization procedure. An interesting approach is the $\varepsilon-$contamined normal neighborhood model of Huber \cite{H64} which was for instance used for Kalman filter purpose in \cite{CR97,SM94}.
However, they consider that the observed variable $Y$ is the hidden state $X$ plus a $\varepsilon-$contamined Gaussian, that is $Y= X + Z\epsilon + (1-Z) O$, instead of Equation~\eqref{eq:modelintro} which better fits to impulse-noised observation modelling. Of note,  some authors \cite{SM94,DK99} assign a cost/weight on each data points during estimation to accommodate from the noise in a similar spirit as the approach we developed in this paper. Unlike our approach, the cost/weight are calculated through optimization problem, making them time consuming.   



In the present paper, we take profit of these modelling assumptions and algorithms by proposing a new model for filtering (and smoothing) sequential data with a possible large number of outliers that may be far from the main pattern. Our modelling is very simple and close to the original Kalman filter and, up to an approximation, relatively inexpensive computationally. It applies for linear Gaussian dynamical system $(X_k)_{k\geq 1}$ and observations $Y_k$ that are distributed as a mixture of Gaussian noised value and independent impulse-noised outliers. We show that the law of the hidden state conditionally to the observation is distributed as a mixture of Gaussian laws. Consequently, we can easily affect a weight for each data and estimate the hidden state by simple recursive argument (without any optimization procedure) based on a law conjugation argument and Kalman filter recursive equation, enabling to have simple explicit formulas.

To our knowledge, even simple, this model and the associated algorithm were never proposed in the literature  while it has a strong potential.

We apply this method to a practical case study. The latter is based on the WoW weighing system detailed in the articles \cite{LLBGGMB21,GANPBALHGP,Wow18,Go18}. This corresponds to a weighing system for small ruminants for which a large number of impulse outliers arise naturally. For this particular case, we also develop an EM-algorithm for which the $E$ part is analytic (no need of Monte-Carlo simulation) and the $M$ part is explicit (no need of optimization algorithm).

\ 

\textbf{Outline:} The paper is now organised on three new sections. Section~\ref{se:kalman} recalls the Kalman filter setting and introduces its generalisation to impulse noised outliers. In Section~\ref{se:cas}, a particular case study is detailed. On this particular case, we develop an EM-algorithm, we test the method on synthetic data with others more classical algorithms and we finally expose our results on real data. We end the paper by a discussion section.





\section{The Kalman Filter with Impulse Noised Outliers (KFINO) algorithm}
\label{se:kalman}

In this section, we recall the definition of the original Kalman Filter in Subsection~\ref{se:KF} and we propose its extension, called Kfino (for Kalman Filter with Impulse Noised Outliers), in Subsection~\ref{se:Kfino}. We end this subsection by simple formula for estimators of interest.

\subsection{Original Kalman Filter}
\label{se:KF}

The basic Kalman filter is a stochastic process which recursively estimates the state of a hidden dynamics system $(X_k)_{k\geq 1}$ in the presence of noised measurement $(Y_k)_{k\geq 1}$. It is a linear dynamical system where the noise is supposed to be Gaussian. Namely $X_1 \sim \mathcal{N}(\mu_1, \Sigma_1)$,
$$
Y_{1} \ | \ X_{1} \sim \mathcal{N} \left( C_1 X_{1}+ d_1, R_1  \right),
$$
and for $k\geq 2$,
\begin{align}
X_{k} \ &| \ X_{k-1} \sim \mathcal{N} \left( A_k X_{k-1} + b_k, Q_k  \right),\label{eq:iter-dis}\\
Y_{k} \ &| \ X_{k} \sim \mathcal{N} \left( C_k X_{k}+ d_k, R_k  \right), \label{eq:iter-dis-y}
\end{align}
where  $\mathcal{N} \left(\mu, \Sigma  \right)$ denotes classically a multivariate Gaussian distribution with mean $\mu$ and co-variance matrix $\Sigma$.  Parameters of the model $\Theta = \{ \mu_1, \Sigma_1, (A_k,Q_k, b_k, d_k, C_k, R_k) , k\geq 1\}$ are the two parameters of the initial condition $X_1$ and a sequence whose terms are two real matrices $A_k,Q_k$, two real vectors $b_k,d_k$ and two co-variance matrices 
$C_k, R_k$.

\subsubsection{Filtering}

The Kalman Filter is a simple but powerful expression of the conditional distribution of $X_k$ given $Y_1, \dots Y_k$ obtained by a recursive treatment along the time (it is a \textit{forward algorithm}). Indeed, it is easy to see that this law is normal with mean $\forward{\mu}_k$ and covariance $\forward{\Sigma}_k$ (the arrow notation indicates the forward or backward computations) which can be computed iteratively by the following two steps: for all $k\geq 1$, 

\noindent \textbf{Propagation step} \\
In this step, we build two intermediate computational mean and covariance: $\forward{\mu}_k^-, \forward{\Sigma}_k^-$. If $k=1$ then we set $\forward{\mu}^-_1 = \mu_1$ and $\forward{\Sigma}^-_1=\Sigma_1$. Otherwise, we set
\begin{align*}
\forward{\mu}_k^-&= A_k \forward{\mu}_{k-1} + b_k, \\
\forward{\Sigma}_k^-&=A_k \forward{\Sigma}_{k-1} A_k^t + Q_k,
\end{align*}
where matrix $A^t$ denotes the transpose of a matrix $A$. This step corresponds to the calculations of the distributions of $X_{k}$ in function of those of $X_{k-1}$ without using the information of $Y_k$. This is directly given by using the iterative relation \eqref{eq:iter-dis}.

\noindent \textbf{Update step}\\
We set
\begin{align*}
\forward{\mu}_k&= \forward{\mu}_k^- +  \left(\forward{\Sigma}_k^- C_k^t \left(C_k \forward{\Sigma}_k^- C_k^t + R_k \right)^{-1} \right) \left( Y_k - (C_k \forward{\mu}_k^-  + d_k ) \right), \\
\forward{\Sigma}_k&= \left( I - \forward{\Sigma}_k^- C_k^t \left(C_k \forward{\Sigma}_k^- C_k^t + R_k \right)^{-1} C_k \right) \forward{\Sigma}_k^-. 
\end{align*}

This corresponds to a correction step by using the measured observation $Y_k$. It is based on the calculations of the conditional distribution of $X_k$ given $Y_k$ using \eqref{eq:iter-dis-y}.

The Kalman filter then consists in iterating these two steps according to the number $N$ of observations. Consequently the number of the computation is of order $N$.

\subsubsection{Smoothing}
\label{se:smoothing}
In addition to the law of $X_k$ conditionally on $Y_1,\dots, Y_k$ derived by the previous \textit{forward algorithm}, we can classically also compute the law of $X_k$ conditionally on $Y_1,\dots, Y_N$, for $k\leq N$ using a \textit{backward algorithm}. This is done by the classical Rauch-Tung-Striebel algorithm \cite{RTS65}. More precisely, the law of  $X_k$ conditionally on $Y_1,\dots, Y_N$, for $k\leq N$ is a normal distribution with
mean $\backward{\mu}_{k}$ and variance matrix $\backward{\Sigma}_{k}$ where these quantities can be calculated recursively as follow : we have
$$
\backward{\mu}_{N}=\forward{\mu}_{N}, \qquad \backward{\Sigma}_{N}= \forward{\Sigma}_{N},
$$
and then for $1\leq k\leq N$, we have
\begin{align*}
K_{k+1}&= \forward{\Sigma}_k A_{k+1}^t (\forward{\Sigma}_{k+1}^-)^{-1} \\
\backward{\mu}_{k}&=\forward{\mu}_k + K_{k+1}(\backward{\mu}_{k+1}-(A_{k+1 }\forward{\mu}_k+b_{k+1}))\\
\backward{\Sigma}_{k} &=  \forward{\Sigma}_k +  K_{k+1} \left( \backward{\Sigma}_{k+1} -\forward{\Sigma}_{k+1}^- \right) K_{k+1}^t.
\end{align*}

\subsection{Kalman filter with impulse noised outliers }
\label{se:Kfino}
In the Kalman filter with impulse noised outliers, we consider three random sequences that are : the hidden state of the system $(X_k)_{k\geq1}$, the observation state of the system $(Y_k)_{k\geq1}$, and a new hidden sequence $(Z_k)_{k\geq 1}$  on $\{0,1\}$ which  encodes if a state is an outlier or not.  We assume that this new sequence is independent from $(X_k)_{k\geq1}$; this translates into assuming that the current state is well monitored independently on its value. 
We further assume that outliers are independent from the dynamics. The dependence scheme is described in Figure~\ref{fig:DAG}.

This gives the following model: $X_1 \sim \mathcal{N}(\mu_1, \Sigma_1)$, $Z_1 \sim \mathcal{B}(p_1)$, where $\mathcal{B}(p)$ denotes classically the Bernoulli distribution with parameter $p \in [0,1]$, 
\begin{align*}
Y_{1} \ &| \ \left( X_{1}, Z_1 =0 \right) \sim \mathcal{P}_1, \\
Y_{1} \ &| \ \left( X_{1}, Z_1 =1 \right) \sim \mathcal{N} \left( C_1 X_{1}+ d_1, R_1  \right),
\end{align*}
and for $k\geq 2$,
\begin{align*}
X_{k} \ &| \ X_{k-1} \sim \mathcal{N} \left( A_k X_{k-1} + b_k, Q_k  \right),\label{eq:iter-dis-O}\\
Z_k& \ \sim \mathcal{B}(p_k) \\
Y_{k} \ &| \ \left( X_{k}, Z_k =0 \right) \sim \mathcal{P}_k, \\
Y_{k} \ &| \ \left( X_{k}, Z_k =1 \right) \sim  \mathcal{N} \left( C_k X_{k}+ d_k, R_k  \right).
\end{align*}

Parameters are now $\Theta = \{ \mu_1, \Sigma_1, (A_k,Q_k, b_k, d_k, C_k, R_k, p_k, \mathcal{P}_k) , k\geq 1\}$, where, in addition to the classical parameters of the Kalman filter, we have a sequence of numbers $(p_k)_{k\geq 1}$ on $[0,1]$ and a sequence of law $(\mathcal{P}_k)_{k\geq 1}$ . We assume that, for every $k\geq 1$, the law $\mathcal{P}_k$ admits a density $\varphi_k$ with respect to the Lebesgue density. We assume of course that the size of the vector $Y_k$ is the same if $Y_k \in \mathcal{N} \left( C_k X_{k}+ d_k, R_k  \right)$ as if $Y_k \in  \mathcal{P}_k$; otherwise it is easy to identify the outliers. In the same way, we denote by $\phi_{\mu, \Sigma}$ the Gaussian density of mean $\mu$ and covariance $\Sigma$.

\begin{center}
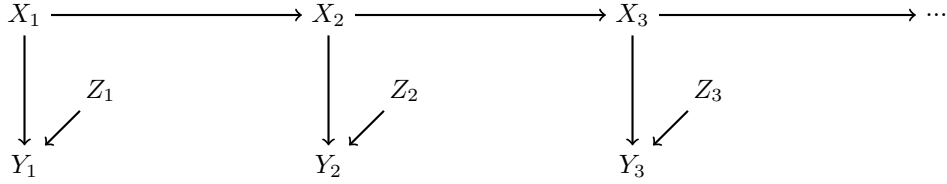
\begin{figure}
  \begin{tikzpicture}
    \node (1) at (0,0) {$X_1$};
    \node (2)  at (4,0) {$X_2$};
    \node (3)  at (8,0) {$X_3$};
    \node (4)  at (12,0) {$...$};
    \node (Z1) at (1,-1) {$Z_1$};
    \node (Z2)  at (5,-1) {$Z_2$};
    \node (Z3)  at (9,-1) {$Z_3$};
    \node (Y1) at (0,-2) {$Y_1$};
    \node (Y2)  at (4,-2) {$Y_2$};
    \node (Y3)  at (8,-2) {$Y_3$};
    \draw [->] [thick] (2) -- (3)  ;
    \draw [->]  [thick]  (1) -- (2);
       \draw [->]  [thick]  (3) -- (4);
    \draw [->] [thick] (2) -- (Y2)  ;
    \draw [->]  [thick]  (1) -- (Y1);
       \draw [->]  [thick]  (3) -- (Y3);
    \draw [->] [thick] (Z2) -- (Y2)  ;
    \draw [->]  [thick]  (Z1) -- (Y1);
       \draw [->]  [thick]  (Z3) -- (Y3);
  \end{tikzpicture}
  \caption{Directed acyclic graph representing the various dependence between $(X_k), (Y_k)$ and $(Z_k)$ }
  \label{fig:DAG}
  \end{figure}
 \end{center}

Simple calculations, based on Bayes formula and iterative argument show that the law of $X_k$ conditionally on $Y_1,..., Y_k$ is a mixture of normal distribution. Indeed, if we condition on $(Z_1, ...., Z_k) = z\in \{0,1\}^k$ then we are (almost) in the classical setting of Kalman Filter. Indeed, conditionally on $Z_k=0$, then $Y_k$ is independent from $X_k$ and then we do not need to do the update step. For $k\geq 1$ and $z\in\mathbb{U}_k =\{0,1\}^k$, let us set $ \forward{\mu_z}$ and $\forward{\Sigma_z}$ be respectively the mean and covariance of $X_k$ conditionally on $Y_1,..., Y_k$ and $(Z_1, ...., Z_k) = z$.

Finally, as expected,  the weights of the mixture, which are the probabilities
$$
\forward{p_z} = \mathbb{P}\left(Z_1=z_1, \dots, Z_k=z_k \ | \ Y_1, \dots, Y_k \right),
$$
for $z=(z_1,\dots, z_k) \in \{0,1\}^k$, can also be computed explicitly. These calculations are detailed in Algorithm~\ref{alg:Kafino-exact}.


\begin{algorithm}
\caption{Exact Kalman filter with impulse noised outliers}\label{alg:Kafino-exact}
\begin{algorithmic}
\Require 
the observation $(y_1,...,y_N)$ with $N\geq 1$ \\  
parameter of the model $\Theta = \{ \mu_1, \Sigma_1, (A_k,Q_k, b_k, d_k, C_k, R_k, p_k, \mathcal{P}_k) , k\geq 1\}$
\Ensure parameters of the conditionnal law of $(X_1,\dots, X_k)$ :  $(\forward {p_z})_{z\in \mathbb{U}_N}$, $(\forward{\mu_z})_{z\in \mathbb{U}_N}$, $(\forward{\Sigma_z})_{z\in \mathbb{U}_N}$\\
likelihood of the observations $(y_1,...,y_N)$ : $L$ 
\State $\ell_0 \gets \phi_{C_1 \mu_1 +d_1, C_1 \Sigma_1 C_1^{t} +R_1} (y_1)$
\State $\ell_1 \gets \varphi_{1} (y_1)$
\State $\forward{p_1}\gets \dfrac{p_1 \ell_1 }{ p_1 \ell_1+ (1-p_1) \ell_0}$
\State $\forward{p_0} \gets \dfrac{(1-p_1) \ell_0 }{ p_1 \ell_1+ (1-p_1) \ell_0}$ 
\State $(\forward{\mu_0},\forward{\Sigma_0}) \gets (\mu_1,\Sigma_1)$ \Comment{This is the propagation step} 
\State $\forward{\mu_1} \gets \forward{\mu_0} +  (\forward{\Sigma_0} C_1^t (C_1 \forward{\Sigma_0} C_1^t + R_1 )^{-1} ) ( y_1 - (C_1 \forward{\mu_0}  + d_1 ) ) $ \Comment{This is the correction step} 
\State $\forward{\Sigma_1} \gets  (I - \forward{\Sigma_0} C_1^t (C_1 \forward{\Sigma_0} C_1^t + R_1 )^{-1} C_1 ) \forward{\Sigma_0} $
\For{$k\in \{2,...,N\}$}
\For{$z\in \{0,1\}^{k-1}$}
\State $\ell_{(z,0)} \gets \ell_z \times \phi_{C_k \forward{\mu_z} +d_k, C_k \forward{\Sigma_z} C_k^{t} +R_k} (y_k)$
\State $\ell_{(z,1)} \gets \ell_z \times \varphi_{k} (y_k)$
\State $\forward{p}_{(z,1)}^- \gets \forward{p_z} p_k  \phi_{C_k \forward{\mu_z} +d_k, C_k \forward{\Sigma}_z C_k^{t} +R_k} (y_k) $\Comment{(temporary unnormalized probabilities)}
\State $\forward{p}_{(z,0)}^- \gets \forward{p_z} (1-p_k) \varphi_{k} (y_k)$%
\State $(\forward{\mu}_{(z,0)},\forward{\Sigma}_{(z,0)}) \gets \left(A_k \forward{\mu}_{z} + b_k , A_k \forward{\Sigma}_{z} A_k^t + Q_k\right)$ \Comment{This is the propagation step} 
\State $\forward{\mu}_{(z,1)} \gets \forward{\mu}_{(z,0)} +  (\forward{\Sigma}_{(z,0)} C_k^t (C_k \forward{\Sigma}_z^- C_k^t + R_k )^{-1} ) ( y_k - (C_k \forward{\mu}_{(z,0)}  + d_k ) ) $ \Comment{This is the correction step} 
\State $\forward{\Sigma}_{(z,1)} \gets  (I - \forward{\Sigma}_{(z,0)} C_k^t (C_k \forward{\Sigma}_{(z,0)} C_k^t + R_k )^{-1} C_k ) \forward{\Sigma}_{(z,0)} $
\EndFor
\State $C \gets \textbf{sum}(\forward{p}_{z}^-, z\in \{0,1\}^{k})$  \Comment{(normalization)}
\For{$z\in \{0,1\}^{k}$}
\State $\forward{p_z} \gets \forward{p}_{z}^- /C$
\EndFor
\EndFor
\State $ L \gets \textbf{sum}(\forward{p}_{z} \ell_z, z\in \{0,1\}^{N})$
\end{algorithmic}
\end{algorithm}

 For a vector $z\in \mathbb{U}$, we set $(z,0)$ and $(z,1)$ for the vectors $(z_1,...,z_k,0)$ and $(z_1,...,z_k,1)$. In addition to all the previous quantities which are helpful to filter the data, we can also, as for the usual Kalman filter, calculate the likelihood $L$ of the observation, which is itself helpful for the estimation of $\Theta$. Here, we have 
 \begin{equation}
 \label{eq:likelihood}
 L = \sum_{z\in \mathbb{U}_N} \left( p^{\sum_{i=1}^N z_i} (1-p)^{N-\sum_{i=1}^N z_i} \right) \ell_z,
 \end{equation}
 where $\ell_z$ are the likelihood of $(Y_1,...,Y_N)$ conditionally on $(Z_1, ...., Z_N) = z$. They are calculated iteratively from the likelihood $\ell_z$ of $(Y_1,...,Y_k)$ conditionally on $(Z_1, ...., Z_k) = z \in \mathbb{U}_k$ for $k\leq N$.
 
 Of course, Algorithm~\ref{alg:Kafino-exact} is longer than algorithm described in Section~\ref{se:KF} but it does not rely on additional steps (such as an optimization step for instance). 
 
 Algorithm~\ref{alg:Kafino-exact} is reminiscent of \textit{switching Kalman filter} (SKF). This algorithm was introduced in 1998 by K. Murphy \cite{M98,M02}. Similarly to the previously introduced references, SKF aims to perform a robust Kalman filter algorithm by considering non Gaussian noise. Instead of considering contaminated normal or Student's t distribution, they consider that observations are normal conditionally to another discrete random variable. It enables to consider different variances at each step. With this different modelling, they also obtain a Gaussian mixture sequence whose the computations are similar to our case. SKF has since been widely used in different contexts and has therefore proven its effectiveness in terms of precision and speed of calculation time; see for instance \cite{WBMGBD04,SMS10,RL08, NG18} among several hundred others. The algorithm we propose then possesses the advantages of SKF (simplicity, robustness, rapidity...) but answers a different question because it is concerned with impulse outliers instead of additive ones.
 
 As for the SKF algorithm, its main drawback is that the number of computations increases exponentially. In contrast with other problems, such as finite hidden Markov chain for which the Baum algorithm \cite{BP96,MDZ97} reduces this exponential complexity into a linear one, here it is impossible to reduce it with an exact algorithm because the conditional law is exactly a Gaussian mixture model with an exponential number of components. To overcome this problem, \cite{M98} proposes several approaches. After a number of $\kappa$ possible iterations,  which ensure some vectors $(p_z)_{z\in\{ 0,1\}^\kappa}$,  $(\ell_z)_{z\in\{ 0,1\}^\kappa}$,  $(\forward{\mu_z})_{z\in\{ 0,1\}^\kappa}$, $(\forward{\Sigma_z})_{z\in\{ 0,1\}^\kappa}$. The next iteration provides an update of these vectors of size $2\kappa$. However, to stop the exponential growth of the number of calculations, it is necessary to keep constant the size of the vectors that we iterate. The possible issues to reduce these vectors of size $2\kappa$ into vectors of size $\kappa$ are to collapse them or to select them. Collapsing consists in approximating a mixture of $\kappa^2$ Gaussian into a mixture of $\kappa$ Gaussian. This makes sense and is one of the most used approach; see for instance \cite{M98,NG18,WBMGBD04,M02}. However, this adds additional  computations in relation to a straightforward selection. The other approach simply consists to keep the higher-probability paths. It can be seen as a Multiple Hypothesis Tracking procedure \cite{BF90,KLCR15} or is similar to the selection procedure of the now well-used SMC algorithms \cite{CMT07,DM13}. 
 See \cite{BK13} for an heuristic but well argued approach which supports the efficiency of such truncation. 
After this truncation step, we can update our parameters as before and then use a new truncation step and so on. This leads to Algorithm~\ref{alg:Kafino}.

\begin{algorithm}
\caption{Kalman filter with impulse noised outliers (Kfino)}\label{alg:Kafino}
\begin{algorithmic}
\Require 
the observation $(y_1,...,y_N)$ with $N\geq 1$ \\  
parameter of the model $\Theta = \{ \mu_1, \Sigma_1, (A_k,Q_k, b_k, d_k, C_k, R_k, p_k, \mathcal{P}_k) , k\geq 1\}$\\
the truncation parameter $\kappa < N$
\Ensure parameters of the conditional law of $(X_1,\dots, X_k)$ :  $(\forward {p_z})_{z\in \mathbb{U}_N}$, $(\forward{\mu_z})_{z\in \mathbb{U}_N}$, $(\forward{\Sigma_z})_{z\in \mathbb{U}_N}$\\
likelihood of the observations $(y_1,...,y_N)$ : $L$ 
\State $\ell_0 \gets \phi_{C_1 \mu_1 +d_1, C_1 \Sigma_1 C_1^{t} +R_1} (y_1)$
\State $\ell_1 \gets \varphi_{1} (y_1)$
\State $\forward{p_1} \gets \dfrac{p_1 \ell_1 }{ p_1 \ell_1+ (1-p_1) \ell_0}$
\State $\forward{p_0} \gets \dfrac{(1-p_1) \ell_0 }{ p_1 \ell_1+ (1-p_1) \ell_0}$ 
\State $(\forward{\mu_0},\forward{\Sigma_0}) \gets (\mu_1,\Sigma_1)$ \Comment{Propagation step} 
\State $\forward{\mu_1} \gets \forward{\mu_0} +  (\forward{\Sigma_0} C_1^t (C_1 \forward{\Sigma_0} C_1^t + R_1 )^{-1} ) ( y_1 - (C_1 \forward{\mu_0}  + d_1 ) ) $ \Comment{Correction step} 
\State $\forward{\Sigma_1} \gets  (I - \forward{\Sigma_0} C_1^t (C_1 \forward{\Sigma_0} C_1^t + R_1 )^{-1} C_1 ) \forward{\Sigma_0} $
\For{$k\in \{2,...,\kappa\}$}
\For{$z\in \{0,1\}^{k-1}$}
\State $\ell_{(z,0)} \gets \ell_z \times \phi_{C_k \forward{\mu_z} +d_k, C_k \forward{\Sigma_z} C_k^{t} +R_k} (y_k)$
\State $\ell_{(z,1)} \gets \ell_z \times \varphi_{k} (y_k)$
\State $\forward{p}_{(z,1)}^- \gets \forward{p_z} p_k  \phi_{C_k \forward{\mu_z} +d_k, C_k \forward{\Sigma}_z C_k^{t} +R_k} (y_k) $\Comment{(temporary unnormalized probabilities)}
\State $\forward{p}_{(z,0)}^- \gets \forward{p_z} (1-p_k) \varphi_{k} (y_k)$%
\State $(\forward{\mu}_{(z,0)},\forward{\Sigma}_{(z,0)}) \gets \left(A_k \forward{\mu}_{z} + b_k , A_k \forward{\Sigma}_{z} A_k^t + Q_k\right)$ \Comment{Propagation step} 
\State $\forward{\mu}_{(z,1)} \gets \forward{\mu}_{(z,0)} +  (\forward{\Sigma}_{(z,0)} C_k^t (C_k \forward{\Sigma}_z^- C_k^t + R_k )^{-1} ) ( y_k - (C_k \forward{\mu}_{(z,0)}  + d_k ) ) $ \Comment{Correction step} 
\State $\forward{\Sigma}_{(z,1)} \gets  (I - \forward{\Sigma}_{(z,0)} C_k^t (C_k \forward{\Sigma}_{(z,0)} C_k^t + R_k )^{-1} C_k ) \forward{\Sigma}_{(z,0)} $
\EndFor
\State $C \gets \textbf{sum}(\forward{p}_{z}^-, z\in \{0,1\}^{k})$  \Comment{(normalization)}
\For{$z\in \{0,1\}^{k}$}
\State $\forward{p_z} \gets \forward{p}_{z}^- /C$
\EndFor
\EndFor
\State $ \mathcal{I} \gets \{0,1\}^{\kappa}$ \Comment{Selected index (trunctation)}
\For{$k\in \{\kappa+1,...,N\}$}
\For{$z\in \mathcal{I}$}
\State $\ell_{(z,0)} \gets \ell_z \times \phi_{C_k \forward{\mu_z} +d_k, C_k \forward{\Sigma_z} C_k^{t} +R_k} (y_k)$
\State $\ell_{(z,1)} \gets \ell_z \times \varphi_{k} (y_k)$
\State $\forward{p}_{(z,1)}^- \gets \forward{p_z} p_k  \phi_{C_k \forward{\mu_z} +d_k, C_k \forward{\Sigma}_z C_k^{t} +R_k} (y_k) $\Comment{(temporary unnormalized probabilities)}
\State $\forward{p}_{(z,0)}^- \gets \forward{p_z} (1-p_k) \varphi_{k} (y_k)$%
\State $(\forward{\mu}_{(z,0)},\forward{\Sigma}_{(z,0)}) \gets \left(A_k \forward{\mu}_{z} + b_k , A_k \forward{\Sigma}_{z} A_k^t + Q_k\right)$ \Comment{Propagation step} 
\State $\forward{\mu}_{(z,1)} \gets \forward{\mu}_{(z,0)} +  (\forward{\Sigma}_{(z,0)} C_k^t (C_k \forward{\Sigma}_z^- C_k^t + R_k )^{-1} ) ( y_k - (C_k \forward{\mu}_{(z,0)}  + d_k ) ) $ \Comment{Correction step} 
\State $\forward{\Sigma}_{(z,1)} \gets  (I - \forward{\Sigma}_{(z,0)} C_k^t (C_k \forward{\Sigma}_{(z,0)} C_k^t + R_k )^{-1} C_k ) \forward{\Sigma}_{(z,0)} $
\EndFor
\State $C \gets \textbf{sum}(\forward{p}_{z}^-, z\in \{0,1\}^{k})$  \Comment{(normalization)}
\For{$z\in \{0,1\}^{k}$}
\State $\forward{p_z} \gets \forward{p}_{z}^- /C$
\EndFor
\State $ \mathcal{I} \gets$ the $\kappa$ first indexes of $(\forward{p_z})_{z\in \mathcal{I} \times \{0,1\}}$ sorted in descending order.\Comment{Selected index update}
\EndFor
\State $ L \gets \textbf{sum}(\forward{p}_{z} \ell_z, z\in \{0,1\}^{N})$
\end{algorithmic}
\end{algorithm}

To end this section, let us rapidly evoke the \textit{smoothing}. As can be seen in Subsection~\ref{se:smoothing}, the Rauch-Tung-Striebel algorithm does not involve the observation. To estimate the law of the first state $X_1$ with respect to all the observation data $Y_1, ..., Y_N$, we start from the law of the last state $X_N$ with respect to all the observation data, given by the \textit{filtering} state, and then we go backward using only the Formula~\eqref{eq:iter-dis}. Consequently, adding outlier as previously defined, does not affect the \textit{backward step}. When we end the \textit{filtering} step, we have some families of parameters $(\forward{p_z})_{z\in\mathcal{I}}$,  $(\ell_z)_{z\in\mathcal{I}}$,  $(\forward{\mu_z})_{z\in\mathcal{I}}$, $(\forward{\Sigma_z})_{z\in\mathcal{I}}$ for some subset $\mathcal{I} \subset \{0,1\}^N$ (whose the cardinal is $2^\kappa$). For all $z\in \mathcal{I}$, we can set $\backward{\mu}_{N,z} = \forward{\mu}_{z} $, $\backward{\Sigma}_{N,z}=\forward{\Sigma}_{z}$ and then build $\backward{\mu}_{k,z}$ and $\backward{\Sigma}_{k,z}$ from the previously build $\backward{\mu}_{k,z}$, $\backward{\Sigma}_{k,z}$ through the iteration relation of Subsection~\ref{se:smoothing}. We then have the law of $X_k$ given $Y_1, ..., Y_N$ and $Z_1$ by considering a mixture of Gaussian, with these means and covariance and associated mixture rates $(\forward{p}_z)_{z\in \mathcal{I}}$.

\subsection{\textit{A posteriori} estimators}
\label{se:estimator}
The outputs of Kfino Algorithm naturally allow us to provide several estimators of interest that we quickly reveal here. These latter are used in Section~\ref{se:cas} to evaluate the algorithm on synthetic data or to predict on real ones. In what follows, we used the natural estimators coming from the forward step but they can be easily extended with outputs coming from the additional backward step.

For every $k\in \{1, ..., N\}$, the posterior mean estimator $\widehat{X}_k$ of the hidden state $X_k$ is given
$$
\widehat{X}_k = \mathbb{E}\left[ X_k \ | \ Y_1, \dots, Y_k  \right] = \sum_{z \in \mathbb{U}_k} \forward{p_z} \forward{\mu_z}.
$$
A path of this estimator on real data is represented by the black line in the bottom right subgraph of Figure~\ref{fig:data-brut}.

The posterior mean estimator $\widehat{Z}^{\text{PM}}_k$ of the hidden state $Z_k$ is given
$$
\widehat{Z}^{\text{PM}}_k = \mathbb{E}\left[ Z_k \ | \ Y_1, \dots, Y_k  \right] = \sum_{z \in \mathbb{U}_k} \forward{p_z} \mathbf{1}_{z_k=1}.
$$
It can be understood as the \textit{a posteriori} probability of not being an outlier. Another easily calculable and illustrative estimator is given by the \textit{maximum a posteriori} estimator :
$$
\widehat{Z}^{\text{MAP}}_k = \underset{{z \in \{0,1\}}}{\text{argmax}} \ \mathbb{P}\left( Z_k=z  \ | \ Y_1, \dots, Y_k  \right) = \mathbf{1}_{ \widehat{Z}^{\text{PM}}_k > 0.5 }.
$$
This latter estimator is also represented in the bottom right subgraph of Figure~\ref{fig:data-brut}. Indeed, each data $k$ is represented in violet when $\widehat{Z}^{\text{MAP}}_k =0$ and in black when $\widehat{Z}^{\text{MAP}}_k =1$.

Finally a last useful estimator is given by the  covariance matrix of the posterior distribution  $\widehat{\Sigma_k}$ which is given by
$$
\widehat{\Sigma_k} = \sum_{z \in \mathbb{U}_k} \forward{p_z} \left( \forward{\Sigma_z} +  \forward{\mu_z} (\forward{\mu_z})^t \right) + \left( \sum_{z \in \mathbb{U}_k} \forward{p_z}   \forward{\mu_z} \right) \left( \sum_{z \in \mathbb{U}_k} \forward{p_z}   \forward{\mu_z} \right)^t.
$$
This estimator is used in Figure~\ref{fig:data-brut} to build the confidence interval (namely, we use a Gaussian type approximation by considering a band of larger $\pm 2\widehat{\Sigma}^{1/2}$).

\section{A case study}
\label{se:cas} 
\subsection{Context}
The motivation of the simple new algorithm we present comes from \cite{LLBGGMB21,GANPBALHGP,Wow18,Go18}. In these works, a new bodyweight /  liveweight (BW)  monitoring system for small ruminant is introduced.  This Walk-over-Weighting (WoW) prototype allows to automatically measure weights (without labor consuming of static measurement) at high frequency (more than one measure per days per animal) in indoor or outdoor setting. This technology has a strong potential for livestock management and animal health. However, due to the gregarious behavior of small ruminants, the WoW prototype gives a large number of (impulse noised) outliers. For instance, in \cite{LLBGGMB21}, $26.4\%$ ($1429/5411$ data) were retained after outlier detection in a first phase, and $38.7\%$ in a second phase. Right hand side of Figure~\ref{fig:data-brut} comes from this experiment data set. In a different experimental setting, \cite{Wow18} finds between $60.1\%$ and $68.7\%$ of retained data. This contrasts with works on cattle. For instance,  in \cite{GBHC14}  they find around $12\%$ of extreme outliers and $5\%$ of others outliers.

In this framework, we are not in the classical setting of a small number of additive outliers. To clean the data, the previous works \cite{LLBGGMB21,GANPBALHGP,Wow18,Go18} used a static measurement plus a confidence interval. This permits to exclude a large number of outliers. However, the use of this former measure is incompatible with an automatic recording. To our knowledge, this type of data was never processed for small ruminants without preliminary measurement. In the cattle case, the statistical method to clean the data was based on local regression method \cite{GBHC14}. They used a pre-processing step by removing extreme values. In a second step, the data are fitted to B-splines penalised on the coefficients. On all their dataset, they find around $12\%$ of extreme outliers and $5\%$ of others outliers.  To have an idea, Figure~\ref{fig:data-brut} illustrates the raw data and  different cleaning methods including a local regression.

In what follows, we propose a mechanistic model for the BW evolution by a Gaussian model. Note that even if we have a finite number of measurements, the weight evolves continuously with time and there is no procedure for the observations: animals go to the WoW when they want. Consequently, we need a continuous time model to take into account that the evolution between two measures can differ.

\subsection{Model construction}

We assume that the weight of an animal evolves as an Ornstein-Uhlenbeck process $(\Xi_t)_{t\geq 0}$ in continuous time; see for instance \cite[Section 8.4.1]{lG16}. This model is not intended to be applied over the whole life of the individuals but only over a few months and can be understood as a local linear approximation. More precisely, we assume that it is solution to the stochastic differential equation :
$$
d \Xi_t = -\textbf{a} \left( \Xi_t - \mathbf{m} \right) dt + \sigma_m d B_t,
$$
where $(B_t)$ is a standard Brownian motion. This modelling possesses several advantages. It allows to transcribe different patterns such as growths or stationary behaviors. Denoting by $t_k>0$ (for $k\geq 1$) the time of the $k^{th}$ measurement then the sequence $(X_k)$, with terms $X_k = \Xi_{t_k}$, is a
simple auto-regressive (AR) model. Parameters of this AR naturally take into account that the observations are taken at non-regular time intervals. This AR relation is given by

$$
X_{k} | X_{k-1} \sim \mathcal{N}\left(X_{k-1} e^{-\textbf{a} (t_{k}-t_{k-1})} + \mathbf{m} (1-e^{-\textbf{a} (t_{k}-t_{k-1})}), \frac{\sigma^2_m}{2\textbf{a}} (1-e^{-2\textbf{a} (t_{k}-t_{k-1})}) \right).
$$
In particular, we have
$$
A_k= e^{-\textbf{a} (t_{k}-t_{k-1})}, \quad b_k =  \mathbf{m} (1-e^{-\textbf{a} (t_{k}-t_{k-1})}), \quad Q_k =   \frac{\sigma^2_m}{2\textbf{a}} \left(1-e^{-2\textbf{a} (t_{k}-t_{k-1})}\right). 
$$
We assume that the probability that an animal is correctly weighted is independent of the time. Namely $p_k =\mathbf{p}$ is constant. Under good measurement, the weight is just a noised modification of the correct weight:
$$
Y_{k} | (X_k, Z_k=1) \sim \mathcal{N}(X_k, \sigma_p^2).
$$
This gives
$$
C_k=1, \quad d_k=0, \quad R_k =\sigma_p.
$$
Otherwise, we also assume that $\mathcal{P}_k$ is constant (over time), and then $\varphi_k=\varphi$ is constant (in $k$). An uninformative choice, which does not work too badly, is to take a uniform distribution. However, as $\Xi_0$ is unknown in our setting, it may be difficult to discriminate which alignment of points in the Figure~\ref{fig:data-brut} on the right corresponds to the evolution of the weight (was $\Xi_0$ $27$ or $40$?). 
 On this figure, it seems impossible, without previous knowledge, to decide. To overcome this problem we will use that it is known that the WoW system more often overestimate the weights when it is wrong (it is because of the animals running or pressing too hard on the weighing scale or because they are two on the weighing). We thus naturally choose
 
\begin{equation}
\label{eq:phi}
    \varphi : y \mapsto \frac{2}{6(M_{\text{max}}-M_{\text{min}})} +
 \frac{8 (y - M_{\text{min}})}{6(M_{\text{max}}-M_{\text{min}})^2}.
\end{equation}

This is simply a trapezoidal distribution verifying $\varphi(M_{\text{max}}) =5 \varphi(M_{\text{min}})$ over $[M_{\text{min}},M_{\text{max}}]$. By expertise, we also fix some values that are
$$
\mathbf{a} = 0.001, \ \sigma^2_m= 0.05, \ \sigma^2_p =5.
$$
Namely, $\mathbf{a}$ and $\sigma^2_m$ depend on the time unit and corresponds here to a small impact of $\mathbf{a}$ and the possible evolution of around  $\pm 0.5$kg per day. On the other hand, $\sigma^2_p$ depends on the weight unit and corresponds to acceptable error of the body weight (without being considered as outlier). Parameters $M_{\text{max}}, M_{\text{min}}$ depends on the design experiments, typically the ages and breeds of the herd. We fix $\Sigma_1=1$ for an acceptable error on our estimation of the initial weight. Finally, we consider that $(\mu_1,\mathbf{p}, \mathbf{m})$ are individual parameters that we have to determine. Algorithm~\ref{alg:Kafino} gives as output an estimate of the likelihood that can be maximized. However, we give in Subsection~\ref{se:EM} below, an alternative estimation procedure based on the \textit{Expectation-Maximization} (EM) algorithm which improves the speed and accuracy of our method.

\begin{remark}[Linear computations]
\label{rem:linear}
We can easily see, from Algorithm~\ref{alg:Kafino}, that, in our example, for every $z\in \{0,1\}^k$, we have 
$$
\forward{\mu}_{z} = a_{z} \mu_1 + b_{z} \mathbf{m} + c_z
$$
where $a_z,b_z,c_z$ can be computed by the iterations
$$
a_{(z,0)} = e^{-\mathbf{a}(t_{k+1}-t_k)} a_z, \ b_{(z,0)}= e^{-\mathbf{a}(t_{k+1}-t_k)} b_z + (1-e^{-\mathbf{a}(t_{k+1}-t_k)}) , \ c_{(z,0)} = e^{-\mathbf{a}(t_{k+1}-t_k)} c_z,
$$
and
$$
a_{(z,1)} = \frac{\sigma^2_p}{\forward{\Sigma}_{u0} + \sigma^2_p} a_{(z,0)}, \ b_{(z,1)} = \frac{\sigma^2_p}{\forward{\Sigma}_{u0} + \sigma^2_p} b_{(z,0)} , \ c_{(z,1)} = \frac{\sigma^2_p}{\forward{\Sigma}_{u0} + \sigma^2_p} c_{(z,0)} +  \frac{\forward{\Sigma}_{u0}}{\forward{\Sigma}_{u0} + \sigma^2_p} y_{k+1},
$$
with initial conditions
$$
a_0=1, \ b_0= 0, \ c_0=0, \ a_1= \frac{\sigma^2_p}{\Sigma_1 + \sigma^2_p}, \ b_1= 0 , \ c_1 = \frac{y_1 \Sigma_1}{\Sigma_1 + \sigma^2_p}.
$$
\end{remark}
\subsection{An \textit{Expectation-Maximization} algorithm as estimation procedure}
\label{se:EM}

Maximizing the likelihood in \eqref{eq:likelihood} over $\theta=(\mu_1,\mathbf{p}, \mathbf{m})$ represents different difficulties. For instance,  it is the sum of various very small quantities and is then exposed to numerical approximation. Moreover, in our setting, we have to maximize it over $3$ parameters which multiplies the calculation time by at least several thousand iterations.

From the form of \eqref{eq:likelihood}, a usual  approach is to use an EM algorithm. Indeed, we can consider $L$ as the marginal of the likelihood $L_\theta(Y,Z)$  of $Y=(Y_1,...,Y_N)$ and the Bernoulli sequence $Z=(Z_1,...,Z_N)$. This one reads
$$
L_\theta(Y,Z) = \mathbf{p}^{\sum_{i=1}^N Z_i} (1-\mathbf{p}^{N-\sum_{i=1}^N Z_i}) \ell_z
$$
where $\ell_z=\ell_z(\theta)$ is iteratively constructed through Algorithm~\ref{alg:Kafino} and only depends on $(\mu_1, \mathbf{m})$. The EM algorithm is based on the following procedure : given some value $\theta_k = (\mu^k_1,\mathbf{p}^k, \mathbf{m}^k)  \in (M_{\text{min}},M_{\text{max}}) \times [0,1] \times (M_{\text{min}},M_{\text{max}})$, we then maximize over $\theta$ the following expectation (over $Z$ whose distribution is a Bernoulli/binomial trial with parameters $N$ and $\mathbf{p}^k$):
\begin{align*}
Q(\theta, \theta_k)= \mathbb{E}[\log\left(L_\theta(y,Z)\right) \ | \ Y=y ] 
&= \sum_{z\in \{0,1\}^N} p_z \left( z_i \log(\mathbf{p}) + \left(N - \sum_{i=1}^N z_i\right) \log(1-\mathbf{p}) + \log(\ell_z(\theta)) \right).
\end{align*}
In the previous quantity $ p_z= p_z(\theta_k)$ are obtained by Algorithm~\ref{alg:Kafino} with entries $(\mu^k_1,\mathbf{p}^k, \mathbf{m}^k)$ (and the others fixed values). We set $\theta_{k+1} = \text{argmax } Q(\theta, \theta_k)$ and iterate. It is known that sequence $(\theta_k)$ converges to a local maximum of $L$. In our setting, the maximization procedure can be carried out by hand. Indeed, on the first hand, we have
\begin{equation}
\label{eq:pk1}
\mathbf{p}^{k+1} = \frac{1}{N} \sum_{z\in \{0,1\}^N} p_z \sum_{i=1}^N z_i,
\end{equation}
which corresponds to the mean number of points considered as outliers. On the other hand, we have 
\begin{align*}
\log(\ell_z) &= \log(\varphi(y_1)) \mathbf{1}_{z_1=0} + \log(\phi_{\mu_{1},\Sigma_1 + \sigma^2_p} (y_1)) \mathbf{1}_{z_1=1} \\
&+ \sum_{k=2}^N \left( \log(\varphi(y_k)) \mathbf{1}_{z_k=0} + \log(\phi_{\forward{\mu}_{z_{-k}},\forward{\Sigma}_{z_{-k}} + \sigma^2_p} (y_k))\mathbf{1}_{z_k=1} \right) ,
\end{align*}
where for $z\in \{0,1\}^N$, we used the notation $z_{-i} = (z_1,...,z_{i-1})$. The maximization over $(\mu^k_1, \mathbf{m}^k)$ is then equivalent to the minimization of
\begin{align*}
&\sum_{z\in \{0,1\}^N} p_z \left( \mathbf{1}_{z_1=1} \frac{1}{2}  \frac{\left(y_1 - \mu_{1}\right)^2}{\Sigma_{1} + \sigma^2_p} + \sum_{k=2}^N \mathbf{1}_{z_k=1} \frac{1}{2}  \frac{\left(y_k - \forward{\mu}_{z_{-k}}\right)^2}{\forward{\Sigma}_{z_{-k}} + \sigma^2_p} \right)\\
=& \sum_{z'\in \{0,1\}^{N-1}} p_{1z'}  \frac{1}{2}  \frac{\left(y_1 - \mu_{1}\right)^2}{\Sigma_{1} + \sigma^2_p} + \sum_{k=1}^{N-1} \sum_{z\in \{0,1\}^k}  \frac{1}{2}  \frac{\left(y_{k+1} - \forward{\mu}_{z}\right)^2}{\forward{\Sigma}_{z} + \sigma^2_p} \sum_{z'\in \{0,1\}^{N-k-1}} p_{(z,1,z')}\\
=& \sum_{z'\in \{0,1\}^{N-1}} p_{1z'}  \frac{1}{2}  \frac{\left(y_1 - \mu_{1}\right)^2}{\Sigma_{1} + \sigma^2_p} + \sum_{k=1}^{N-1} \sum_{z\in \{0,1\}^k} \frac{1}{2}  \frac{\left(y_{k+1} - a_z \mu_1 - b_z \mathbf{m} - c_z\right)^2}{\forward{\Sigma}_{z} + \sigma^2_p} \sum_{z'\in \{0,1\}^{N-k-1}} p_{(z,1,z')},
\end{align*}
where in the last line we used Remark~\ref{rem:linear}. Optimizing, we find
\begin{equation}
\label{eq:mk1}
\mu_1^{k+1}= \frac{C Y_b - B Y_a}{C^2-AB}, \ \mathbf{m}^{k+1} = \frac{Y_a C - A Y_b}{C^2 - AB},
\end{equation}
where
\begin{align*}
Y_a&= \sum_{z'\in \{0,1\}^{N-1}} p_{1z'}   \frac{y_1}{\Sigma_{1} + \sigma^2_p} + \sum_{k=1}^{N-1} \sum_{z\in \{0,1\}^k}  a_z \frac{\left(y_{k+1}- c_z\right)}{\forward{\Sigma}_{z} + \sigma^2_p} \sum_{z'\in \{0,1\}^{N-k-1}} p_{(z,1,z')}, \\
A&=\sum_{z'\in \{0,1\}^{N-1}} p_{1z'}   \frac{1}{\Sigma_{1} + \sigma^2_p} + \sum_{k=1}^{N-1} \sum_{z\in \{0,1\}^k}   \frac{ a_z^2 }{\forward{\Sigma}_{z} + \sigma^2_p} \sum_{z'\in \{0,1\}^{N-k-1}} p_{(z,1,z')},\\
C&= \sum_{k=1}^{N-1} \sum_{z\in \{0,1\}^k}   \frac{ a_z b_z }{\forward{\Sigma}_{z} + \sigma^2_p} \sum_{z'\in \{0,1\}^{N-k-1}} p_{(z,1,z')},\\
Y_b&=\sum_{k=1}^{N-1} \sum_{z\in \{0,1\}^k}  \frac{b_z \left(y_{k+1} - c_z\right)}{\forward{\Sigma}_{z} + \sigma^2_p} \sum_{z'\in \{0,1\}^{N-k-1}} p_{(z,1,z')},\\
B&=\sum_{k=1}^{N-1} \sum_{z\in \{0,1\}^k}  \frac{ b_z^2 }{\forward{\Sigma}_{z} + \sigma^2_p} \sum_{z'\in \{0,1\}^{N-k-1}} p_{(z,1,z')}.
\end{align*}

To conclude, to calibrate $\theta$, we start from $\theta_0 \in (M_{\text{min}},M_{\text{max}}) \times [0,1] \times (M_{\text{min}},M_{\text{max}})$, then we use Algorithm~\ref{alg:Kafino} to provide some computational quantities (as $\forward{p}_z$ or $\forward{\Sigma}_z$) and then we update $\theta$ with Equations~\eqref{eq:pk1} and \eqref{eq:mk1}. There is no need for optimization algorithms or stochastic simulations.

\subsection{Numerical tests on synthetic data}

In this section, we test the performance of the Kfino algorithm. We place ourselves in a framework close to the application we are targeting. For this reason we always choose
\begin{equation}
\label{eq:testparam}
 m_0=40, \textbf{m}=60, \ \textbf{a}=0.001, \ M_\text{min}=10, \ M_{\text{max}}=100, \ \sigma^2_{0}=1, \ \sigma^2_{\textbf{m}}=0.05.
\end{equation}
The values of $\mathbf{p}, \sigma_p$ and $\kappa$ depend on the test.

In all numerical tests below, we simulate $100$ different trajectories $((\Xi^i_t)_{t\in [0, 100]})_{i=1,...,100}$ and obtain then $100$ data sets $(Y^i)_{i=1,...,100}$ for which the hidden states are known. Each data set contains $N_i$ points (that are sampled from independent Poisson processes. We then run Algorithm~\ref{alg:Kafino} with known parameters $\Theta$.

For every simulated path $i\in \{1,...,100\}$, we can then calculate the efficiency of Kfino through the following quantities :
\begin{equation}
\label{eq:MSE}
\text{MSE}_i = \frac{1}{N_i} \sqrt{\sum_{k=1}^{N_i} | X^i - \widehat{X}^i |^2}, \qquad A_i = \mathbf{1}_{Z_i} =\hat{Z}^{\text{MAP}_i}
\end{equation}
Variable $\text{MSE}_i$ are the mean square errors and then represent the errors in the filtering and $A_i$ are the accuracy and reflects the quality of the outlier detection.

In Section~\ref{se:trunc}, we test the impact of the value of $\kappa$ on the performance of the algorithm. In others subsection, we compare the Kfino algorithm with two classical algorithms  according to the variation of $\mathbf{p}$ and $\sigma_p$. These two algorithms are a local regression performed by the \textit{locfit} package of R and the second one is the classical Kalman filter. For these two algorithms, a point is considered as outlier when it does not belong to a confidence interval of the reconstructed path on all data.

\begin{figure}
\centering
\includegraphics[scale=0.35]{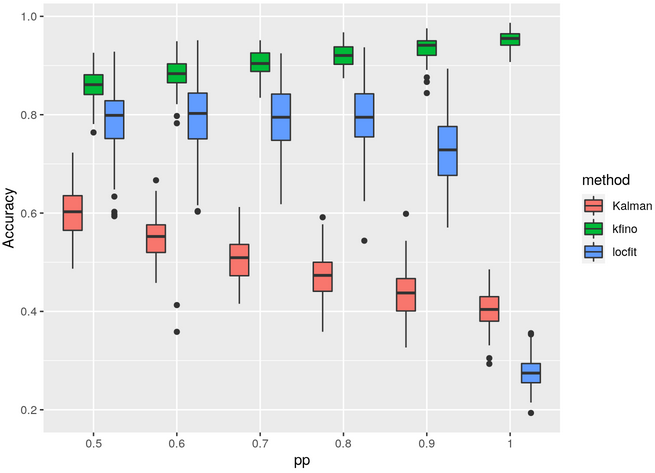}
\includegraphics[scale=0.46]{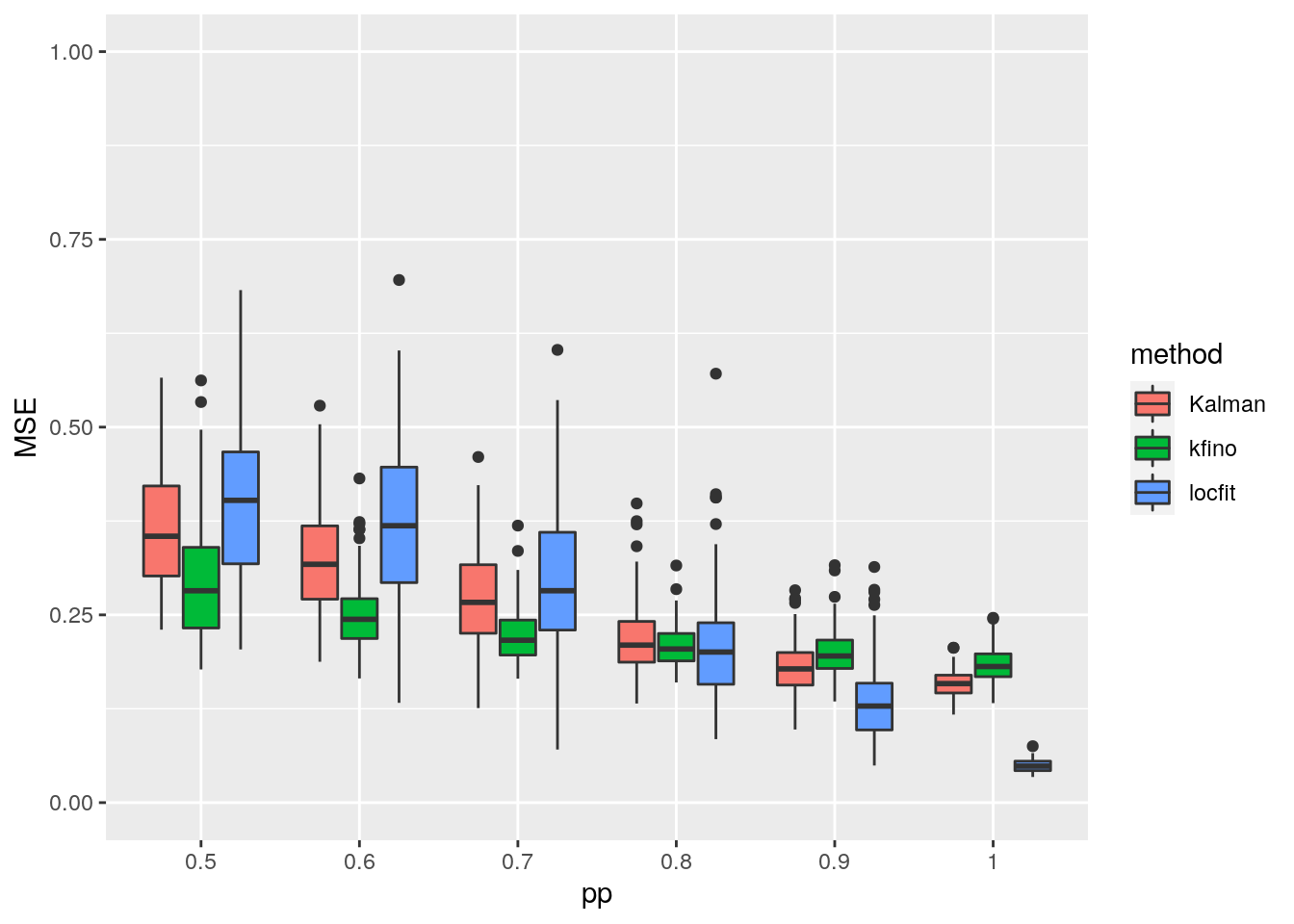}
\caption{Influence of the well-monitored data $p$ on the error and the accuracy. See Section~\ref{se:p} for details.}
\label{fig:bam_p}
\end{figure}

In Figures~\ref{fig:bam_kappa}, \ref{fig:bam_sigma} and \ref{fig:bam_p}, variables $(\text{MSE}_i)$ and $(A_i)$ are represented in box-plots. Let us recall that in box-plots, The middle line indicates the median, the bottom and top edges of the boxes respectively mark the 25th and 75th percentiles, and the maximum and minimum values of the ensemble are marked by the extent of the whiskers.

\subsubsection{Performance of Kfino with respect to the truncation level $\kappa$}
\label{se:trunc}



\begin{figure}
\centering
\includegraphics[scale=0.3]{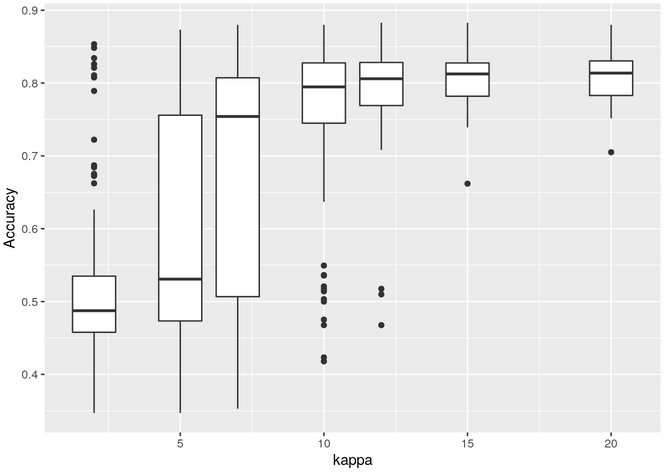}
\includegraphics[scale=0.4]{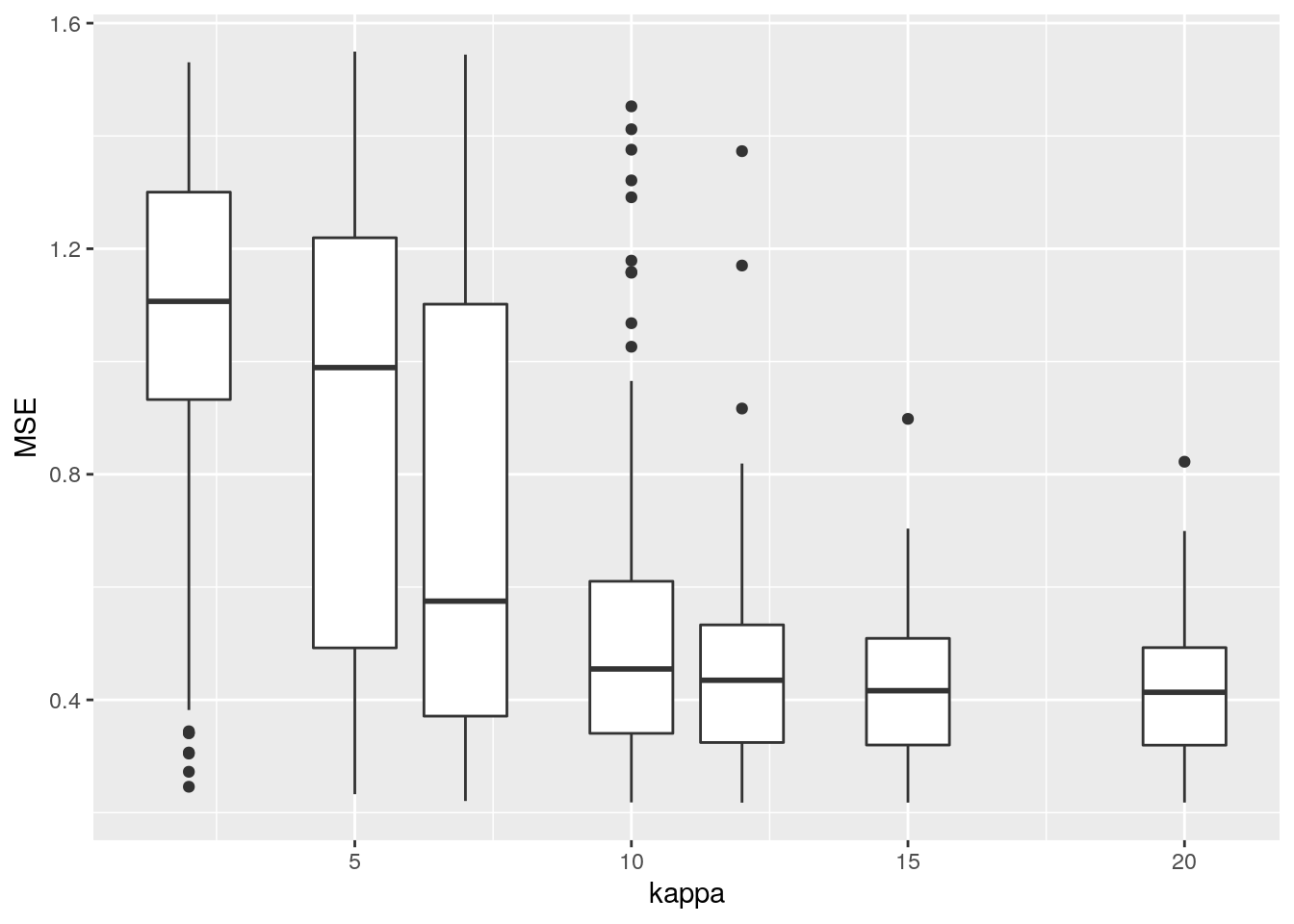}
\caption{Influence of the truncation parameter $\kappa$ on the error and the accuracy. See Section~\ref{se:trunc} for details.}
\label{fig:bam_kappa}

\end{figure}

We tested the performance of the algorithm when the truncation parameter $\kappa$ varies. For each parameter $\kappa \in \{1,5,7,10,12,15,20\}$, we simulated $100$ paths and calculated the error $\text{MSE}_i$ and accuracy $A_i$. We then represented these quantities through box-plots in  Figure~\ref{fig:bam_kappa}. 
We can see in this figure that from $\kappa=7$ at least one simulation over two gives good results and from $\kappa=10$, it is useless to increase $\kappa$ to improve the performance.

Note that in this case, we fix $\textbf{p}=0.5$ and  $\sigma^2_p=5$.

\subsubsection{Performance of Kfino with respect to the level of noise $\sigma_p$}

\label{se:sigmap}
\begin{figure}
\centering
\includegraphics[scale=0.35]{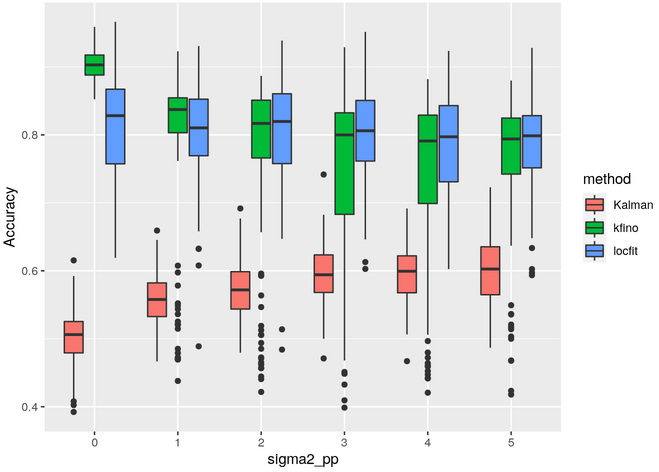}
\includegraphics[scale=0.46]{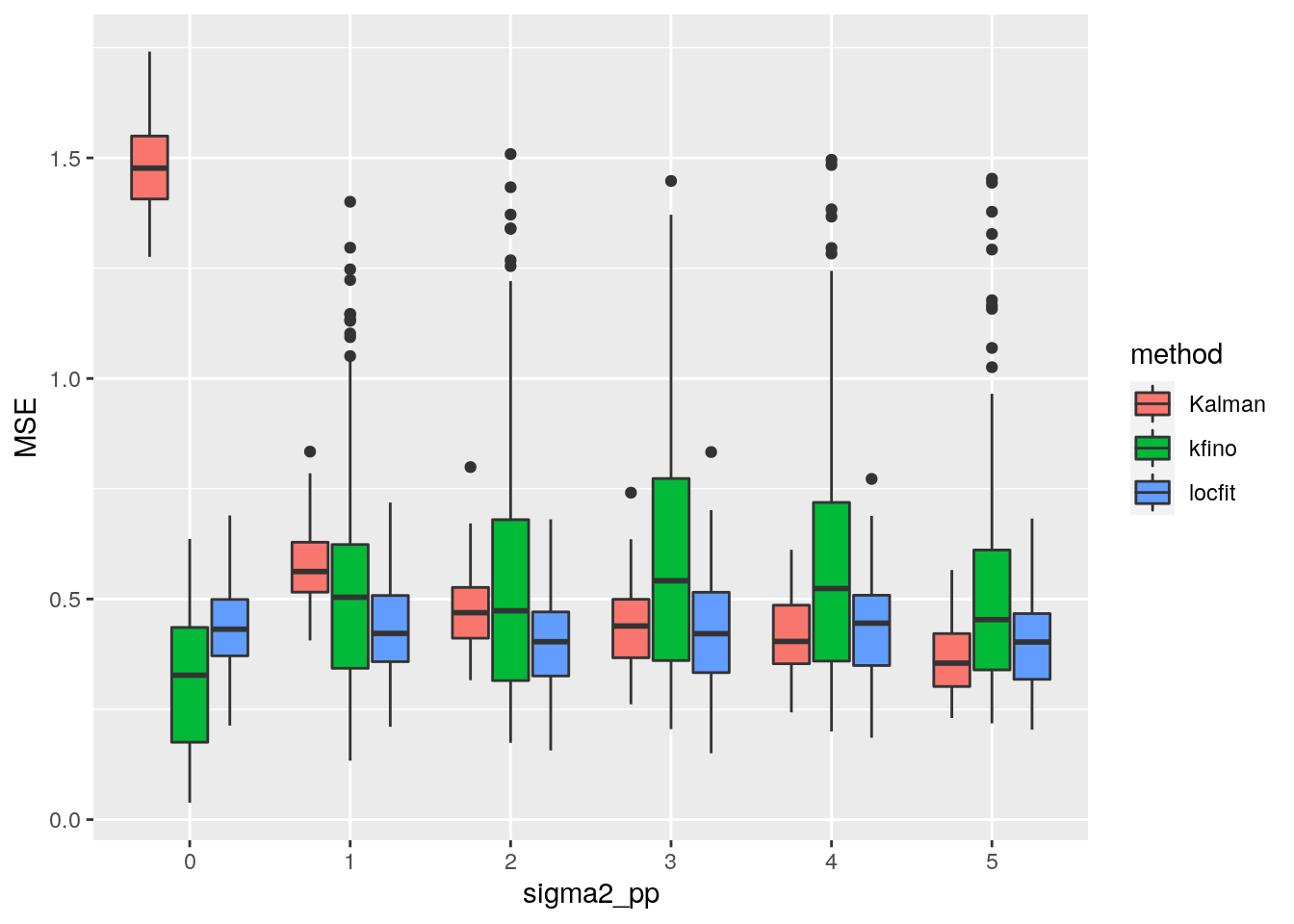}
\caption{Influence of the noise parameter $\sigma_p$ on the error and the accuracy. See Section~\ref{se:sigmap} for details.}
\label{fig:bam_sigma}
\end{figure}

We tested the algorithm when $\sigma_p^2$ varies in $\{0,1,2,3,4,5\}$. In such case, we fixed $\textbf{p}=0.5$ and $\kappa=10$.  An interesting point is that when $\sigma_p=0$ then $Y_k=X_k$ conditionally on $Z_k=1$. In particular, if there was no outlier points then the filtering is useless. We can rightly see in Figure~\ref{fig:bam_sigma} that when $\sigma_p=0$ then the outlier detection is perfect : the algorithm did not fail. When $\sigma_p$ increases then the accuracy converges to $0.8$ which is natural : it is impossible to consider as outlier an outlier $O_k$ which is sufficiently close to $X_k$ even if $Z_k=0$. Both other algorithms have weak performance because they are not adapted for this question. In contrast, we see that Kfino is better then Kalman and local regression for small value of noise $\sigma_p$ but are similar when $\sigma_p$ is large.

\subsubsection{Performance of Kfino with respect to the outlier rate $p$}

\label{se:p}

We tested the algorithm when $p$ varies in $\{0.5,0.6,0.7,0.8,0.9,1\}$. In such case, we fixed $\textbf{p}=0.5$ and $\kappa=10$.  An interesting point is that when $p=1$ then $Z_k=1$ almost-surely and there is no outliers. We can rightly see in Figure~\ref{fig:bam_sigma} that Kfino always have a better accuracy that converges to $1$ when $p=1$. Note that again, both other algorithms have weak performance because they are not adapted for this question. In contrast, we see that Kfino is better then Kalman and local regression for a large number of outliers $p\ll1$ but are similar when $p$ is close to $1$ ($p\geq 0.8$). In particular,  the \textit{locfit} function gives particular goods results when $p=0.9$ or $p=1$ (which goes beyond the framework that motivates us). This is due to two main reasons that are not verified on our real data set. On the first hand, the outlier distribution \eqref{eq:phi} is close to be uniform and outliers are generally not concentrated around the same value; this leads to local regression methods to do not take them into account. On the second hand, the assumption that outliers arise from a Bernoulli scheme gives that outliers are not concentrated in the same time interval. This also lead  local regression methods to do not take them into account.

\subsubsection{Conclusion}

In the setting we present, Kfino algorithm is generally better than the two others one. It always gives better outlier detection. Its performance is particularly better under low noise (small $\sigma$) and large number of outliers (large $p$). Note that \textit{locfit} method seems to be comparable to Kfino algorithm in many cases, but we will see in the next section that it breakdowns when the dataset is less regular (namely, when outliers may be concentrated in time and space). 

\subsection{Real data : small ruminant on the WoW system}


In this section, we will illustrate our method with the experiment \cite{LLBGGMB21}. All individuals were processed and the results are in a supplementary material. Moreover, a companion paper is in preparation to expose in detail the data processing of experiments \cite{LLBGGMB21,GANPBALHGP,Wow18,Go18} with a detailed description of all the results.

To our knowledge, this type of data was never processed (automatically) for small ruminants. Closely related, this type of equipment was already introduced for cattle. In such case, the statistical method to clean the data was based on local regression method \cite{GBHC14}. Similarly to us, they use a pre-processing step by removing extreme values. In a second step, the data are fitted to B-splines penalised on the coefficients. On all their dataset, they find around $12\%$ of extreme outliers (OOR out the range for us) and $5\%$ of others outliers (KO for us). 

\begin{figure}
\centering
\includegraphics[scale=0.55]{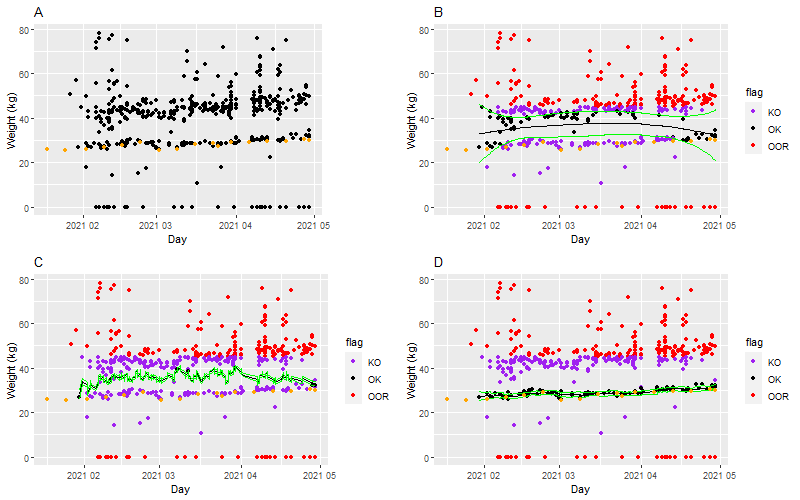}
\caption{Four graphics represent the data analysis of the weighing of the same individual from experiment \cite{LLBGGMB21}. The figure at the top left (A) corresponds to the raw data.  Each black dot represents a weight monitoring. The time is given on the $x-$axis and the measurement on the $y-$axis. The golden dots represent hand measurements that are not used in the algorithm (because they are incompatible with the automation objective) but are given as an indication. In the others figures, red points are data out of the range $[10,45]$, the detected outliers are in purple and the black points are the points considered as well measured. Black line is an estimator of the hidden state and green lines give a confidence interval (given by $\mu$ $\pm$ 2$\sigma$ with $\sigma$ being the estimated square root).
On the top right (B), we used a local regression (using the \textit{locfit} package of R), on the bottom left (C) the classical Kalman filter and on the bottom right (D) the Kfino algorithm.}  
\label{fig:data-brut}
\end{figure}

Nevertheless, because of the gregarious nature of sheep, the number of outliers in our dataset is incomparable to what is observed with the bovine data. We observe around $43.5\%$ of out-of-range data, $21.2\%$ of outlier data. Local regression is not robust enough for such a large number of outliers. Just to illustrate our purpose, see in Figure~\ref{fig:data-brut} an illustration of a local regression type algorithm and of the Kalman filter for a data set on a selected sheep we have to clean.

On this data set consisting in analyzing weighing data on 107 individuals, the Kfino algorithm is able to correctly detect outliers and predict weights in $97\%$ of the cases with sufficient data. These results can be seen in the supplementary material. In addition to the automatic measurement given by the WoW, hand measurements were down in \cite{LLBGGMB21}. This \textit{Gold Standard} (GS) measurement can be considered as the weight of the individual. In contrast with others experiments \cite{LLBGGMB21,GANPBALHGP,Wow18,Go18}, this lead us to easily compare \textit{locfit} and Kfino methods. For each animal, we calculate the obtained mean square error \eqref{eq:MSE} by summing over all GS measurement days and using as $X^i$ this GS measurement. The summary statistics of this MSE are detailled in Table~\ref{tab:rawdata}. The result is indisputable: Kfino gives very good results, whereas locfit has a margin of error that is often too large to be used in practice to detect abnormal weight loss. Note that, for both methods, the maximum MSE is high; this corresponds to three individuals for whom the measurements are too poor to recover an accurate weight estimate.

\begin{table}[]
    \centering
\begin{tabular}{|l|c|c| c|c|c|c|}
  \hline
  Algorithm & Min. & 1st Quart. & Median   & Mean & 3rd Quart. &   Max. \\
  \hline
  \textit{locfit} & 0.6157 & 1.5171 & 1.8710 & 1.9237 & 2.2180 & 4.6101 \\
  Kfino & 0.1124 & 0.2201  & 0.2805 & 0.3960 & 0.3814 & 4.8292 \\
  \hline
\end{tabular}
    \caption{Summary statistics of the MSE of Kfino and \textit{locfit} methods on the real data set of \cite{LLBGGMB21}.  }
    \label{tab:rawdata}
\end{table}
\section{Conclusion/discussion}

To process a dataset with a large number of outliers, we develop the new filtering algorithm Kfino. This one is in line with the classical Kalman Filter but is robust to a large number of outliers. This algorithm is computationally fast because it does not include any optimization algorithm or particle system. It enables to filter the data and to detect the outliers. 

It naturally compares with model-free type algorithms like \textit{locfit} or others local regression algorithms. In this comparison, we recover the usual trade-off between modelling assumptions and adaptability. However, in contrast with others methods, our approach can handle real data set with a large number of outliers badly distributed (in time and in space).

This method is not unrelated to the famous and already used SKF algorithm. Taking into account this remark, several generalization are possibles such that consider a Markovian structure for $(Z_k)_{k\geq1}$ instead of Independence between steps, or consider that $Z_k$ belongs to a larger space than $\{0,1\}$ allowing different type of outliers. In particular, we can consider to treat simultaneously impulse and additive outliers though this algorithm.

The Kfino algorithm shows very good performance on simulated and WoW data. It is adapted for monitoring at random times. We are convinced that it will apply to a more general framework. For this reason, a R package is currently under development.

\ \\
\textbf{Acknowledgement}
This work was financially supported by the European Union’s Horizon 2020 Innovation Action program for funding the project Integrating innovative TECHnologies along the value Chain to improve small ruminant welfARE management (TechCare; grant agreement 862050).

\bibliographystyle{plain}
\bibliography{ref}

\end{document}